\documentclass{article}
\usepackage{graphicx}
\usepackage{amsmath}

\input{tcilatex}

\begin{document}

\bigskip

\bigskip\ \ \ \ \ \ \ \ \ \ \ \ \ \ \ 

\ \ \ \ \ \ \ \ \ \ \ \ \ \ \ \ \ \ \ \ \ \ {\LARGE Simple model of particle
detector and arrival time}

\bigskip

\bigskip

\ \ \ \ \ \ \ \ \ \ \ \ \ \ \ \ \ \ \ \ \ \ \ \ \ \ \ \ \ \ \ \ \ \ \ \ \ \
\ \textbf{\ Viliam Pa\v{z}ma}$^{1}$\textbf{\ and \ J\'{u}lius Vanko}$^{2}$

\bigskip

\ \ \ \ \ \ \ \ \ \ \ \ \ \ \ $^{1}$Department of Theoretical Physic,
Comenius University, Bratislava, Slovakia

\ \ \ \ $^{2}$Department of Nuclear Physics and Biophysics, Comenius
University, Bratislava, Slovakia

\ \ \ \ \ \ \ \ \ \ \ \ \ \ \ \ \ \ \ \ \ \ \ \ \ \ \ \ \ \ \ \ \ \ \ \ \ \
\ \ \ \ \ \ \ \ \ \ \ \ vanko@fmph.uniba.sk

\bigskip

\ \ \ \ \ \ \ \textbf{Abstract \ }: \ We present the very simple model of a
particle detector and the proposal for\ \ \ \ \ \ \ \ \ \ \ \ \ \ \ \ \ \ \
\ \ \ \ \ \ \ \ \ \ \ \ \ \ \ \ \ \ \ \ \ \ \ \ \ \ \ \ \ \ \ \ \ \ \ \ \ \
\ \ \ \ \ \ \ \ \ \ \ \ \ \ \ \ \ \ \ \ \ \ \ \ \ \ \ 

\ \ \ \ \ \ \ \ \ \ \ \ \ \ \ \ \ \ \ \ \ \ \ \ \ \ the calculation \ of the
average value of the time of arrival.

\bigskip

\textbf{1. \ Introduction}

\bigskip

\ \ \ \ In the physics there are questions which have remainded unanswered
for a long time. Besides of others the definitions of the arrival time of
the particles in quantum mechanical sense has belonged among them. There
exist many approaches to this problem but none of them seems to be generally
accepted \ [1,2,3,4].

The close relation between the arrival time and certain model of the
detector is demonstrated in \ [2]. This fact stimulated us to study the
relation between the quantity \ $P_{D}\left( \left\langle
t_{0},t\right\rangle \right) $ , \ which means the probability that a
particle entered the detector during the time interval \ $\left\langle
t_{0},t\right\rangle $ \ and the dynamics of a detector.

In \ [5] \ the detector is taken to be two-state system \ (one state stands
for detection and another for no-detection). This system is coupled to the
environment consisting of large number of oscillators in their ground
states. The coupling between the detector and oscillators is proportional to 
$\Theta \left( x\right) $ . This means that if particle enters the region \ $%
x>0$ \ the detector becomes coupled to the environment. It is shown \ [2] \
that the transition of the detector from no-detection state to detection one
is irreversible (i.e. returning to the no-detection state is not possible)
and the probability of detection at time  $t$ \ is proportional to

\bigskip

\ \ \ \ \ \ \ \ \ \ \ \ \ \ \ \ \ \ \ \ \ \ \ \ \ \ \ \ \ \ \ \ \ \ \ \ \ \
\ \ \ \ \ \ \ $\dint\limits_{0}^{t}dt^{\prime }\dint\limits_{-\infty }^{0}dx$
$\left| \Psi \left( x,t^{\prime }\right) \right| ^{2}$ \ $,$

\bigskip

where \ $\Psi $ \ \ is wave function of considered particle.

Similar models of particle detectors can be found, e.g., in \ [6,7,8].

This paper is organized as follows. In the sec.2 we shall present a very
simple model of particle detector. The expression for probability that
particle entered detector during the time interval \ $\left\langle
t_{0},t\right\rangle $ \ will be proposed in the sec.3. The sec.4 contains
the expression for average value of the arrival time. The sec.5 contains
several notes on problems we studied.

\bigskip

\textbf{2. \ Simple particle detector model}

\bigskip

The model we shall present here is very simple but in our opinion it exibits
irreversible behaviour too. The essence of our considerations consists in
the analysis of following situation.

Let a particle had been emitted at time \ $t_{0}$ \ \ by a point source
placed at \ $\overrightarrow{x}_{0}$ . By this phrase we mean that the wave
function of the particle at time \ $t_{0}$ \ is \ \ $\varphi \left( 
\overrightarrow{x}-\overrightarrow{x}_{0}\right) $ \ \ and moreover we shall
assume that \ $\varphi \neq 0$ \ \ in very small region surrounding the
point \ $\overrightarrow{x}_{0}$ \ \ only. The detector is placed at the
large distance \ $L$ \ \ from \ $\overrightarrow{x}_{0}$ . \ Its volume is \ 
$V_{D}$ \ and defines the spaceangle \ $\Omega _{0}$ \ \ with respect to the
point \ $\overrightarrow{x}_{0}$ .

The detector is taken to be two state system as mentioned above. One state $%
\chi _{0}$ \ stands for no-detection and another \ $\chi _{1}$ \ for
detection. What quantity has forced the detector to the transition from \ $%
\chi _{0}$ \ to \ $\chi _{1}$ \ ? \ First of all, an emitted particle can
enter the detector during time interval \ $\left\langle t_{0},t\right\rangle 
$ \ with the probability \ $P_{D}\left( \left\langle t_{0},t\right\rangle
\right) $ \ (The expression for \ $P_{D}$ \ will be proposed in the sec.3) .
\ The detector will register a particle only if that particle will interact
with it. However, that interaction happen with the probability \ $%
P_{D}\left( \left\langle t_{0},t\right\rangle \right) $ \ only. On the basis
of this we stand for opinion that the ''power'' which has forced the
detector to the transition from \ $\chi _{0}$ \ to \ $\chi _{1}$ \ is the
quantity

\bigskip

\ \ \ \ \ \ \ \ \ \ \ \ \ \ \ \ \ \ \ \ \ \ \ \ \ \ \ \ \ \ \ \ \ \ $\dfrac{d%
}{dt}$ $P_{D}\left( \left\langle t_{0},t\right\rangle \right) $

\bigskip

and the probability of the transition \ $\chi _{0}\rightarrow \chi _{1}$ \
cannot be larger than \ $P_{D}\left( \left\langle t_{0},t\right\rangle
\right) $ . \ Moreover, it is natural to assume that \ $P_{D}\left(
\left\langle t_{0},t\right\rangle \right) $ \ is not decreasing function.

In the next we shall represent states \ $\chi _{0}$ , \ $\chi _{1}$ \ \ by
columns

\bigskip

\ \ \ \ \ \ \ \ \ \ \ \ \ \ \ \ \ \ \ \ \ \ \ 

\ \ \ \ \ \ \ \ \ \ \ \ \ \ \ \ \ \ \ \ \ \ \ \ \ \ \ \ \ \ \ \ \ \ $\chi
_{0}=\left( 
\begin{array}{c}
1 \\ 
0
\end{array}
\right) $ , \ \ \ \ \ \ $\chi _{1}=\left( 
\begin{array}{c}
0 \\ 
1
\end{array}
\right) $

\bigskip

and postulate the equation \ 

\bigskip

\ \ \ \ \ \ \ \ \ \ \ \ \ \ \ \ \ \ \ \ \ \ \ \ \ \ \ \ \ \ \ \ \ \ \ $i%
\dfrac{d}{dt}$ $\chi =H$ $\chi $ $\ \ \ \ \ \ \ \ \ \ \ \ \ \ \ \ \ \ \ \ \
\ \ \ \ \ \ \ \ \ \ \ \ \ \ \ \ \ \ \ \ \ \ \ \ \ \ \ \ \ \ (1)$

\bigskip

with

\ \ \ \ \ \ \ \ \ \ \ \ \ \ \ \ \ \ \ \ \ \ \ \ \ \ \ \ \ \ \ \ \ \ \ $%
H=A\left( t\right) $ $\left( 
\begin{array}{cc}
0 & 1 \\ 
1 & 0
\end{array}
\right) $

\bigskip

as the equation describing the dynamics of the detector \ (Up to now \ $%
A\left( t\right) $ \ is unknown function.) . As the initial state we shall
always choose \ $\chi _{0}$ .

Now from \ (1) \ we get

\bigskip

\ \ \ \ \ \ \ \ \ \ \ \ \ \ \ \ \ \ \ \ \ \ \ \ \ \ \ \ \ \ \ \ \ $\chi
\left( t\right) =\chi _{0}$ $\cos B(t)-i\chi _{1}$ $\sin B(t)$ $,$

\bigskip

where

\ \ \ \ \ \ \ \ \ \ \ \ \ \ \ \ \ \ \ \ \ \ \ \ \ \ \ \ \ \ \ \ \ \ $%
B=\dint\limits_{t_{0}}^{t}dt^{\prime }A\left( t^{\prime }\right) $ $.$

The probability \ \ $P_{reg}$ \ \ of finding the detector in the state \ $%
\chi _{1}$ \ is

\bigskip

\ \ \ \ \ \ \ \ \ \ \ \ \ \ \ \ \ \ \ \ \ \ \ \ \ \ \ \ \ \ \ \ \ \ $%
P_{reg}\left( t\right) =\sin ^{2}B(t)$ $.$ \ \ \ \ \ \ \ \ \ \ \ \ \ \ \ \ \
\ \ \ \ \ \ \ \ \ \ \ \ \ \ \ \ \ \ \ \ \ \ \ \ \ \ \ \ \ \ $(2)$

\bigskip

If the probability that particle entered the detector during \ $\left\langle
t_{0},t\right\rangle $ \ is negligible one can hardly expect that \ $P_{reg}$
\ \ will be large. Moreover, it sems to be natural to assume that

\bigskip

\ \ \ \ \ \ \ \ \ \ \ \ \ \ \ \ \ \ \ \ \ \ \ \ \ \ \ \ \ \ \ \ \ \ $%
P_{reg}\leq P_{D}\left( \left\langle t_{0},t\right\rangle \right) $ $.$

\bigskip

On account of that we postulate the equation

\bigskip

\ \ \ \ \ \ \ \ \ \ \ \ \ \ \ \ \ \ \ \ \ \ \ \ \ \ \ \ \ \ \ \ \ \ $%
P_{reg}=k$ $P_{D}\left( \left\langle t_{0},t\right\rangle \right) $ $,$ \ \
\ \ \ \ \ \ \ \ \ \ \ \ \ \ \ \ \ \ \ \ \ \ \ \ \ \ \ \ \ \ \ \ \ \ \ $(3)$

\bigskip

where \ \ $0<k<1$ .

As one can see \ $A(t)$ \ \ is equal

\bigskip

\ \ \ \ \ \ \ \ \ \ \ \ \ \ \ \ \ \ \ \ \ \ \ \ \ \ \ \ \ \ \ \ \ \ $A\left(
t\right) =\dfrac{d}{dt}\arcsin \sqrt{k\text{ }P_{D}\left( \left\langle
t_{0},t\right\rangle \right) }$ $.$

\bigskip

The requirement \ (3) \ ensures the irreversible behaviour of the detector.

Evidently, this model is too simple. Yet it offers another look on the
problem. Moreover, it inspired us to look for the expression for the
quantity \ $P_{D}\left( \left\langle t_{0},t\right\rangle \right) $ .

\bigskip

\textbf{3. \ The expression for \ \ }$P_{D}\left( \left\langle
t_{0},t\right\rangle \right) $

\bigskip

Let us now consider two events \ $E_{1}$ \ and \ $E_{2}$ . \ The event \ $%
E_{1}$ \ means that emitted particle has momentum \ \ $\overrightarrow{p}\in
\Omega _{D}$ \ (it means that the half-line \ \ $\overrightarrow{x}=%
\overrightarrow{x}_{0}+\overrightarrow{p}$ $s$ \ \ ($s\in \left( 0,\infty
\right) $) \ passes through the detector). \ The event \ $E_{2}$ \ means
that particle in question had been situated in an arbitrary \ \ $t^{\prime
}\in \left\langle t_{0},t\right\rangle $\ \ in the volume \ $V_{D}$ . \ Now
the probability \ $P_{D}\left( \left\langle t_{0},t\right\rangle \right) $ \
can be expressed as

\bigskip

\ \ \ \ \ \ \ \ \ \ \ \ \ \ \ \ \ \ \ \ \ \ \ \ \ \ \ \ \ \ \ \ \ \ \ \ $%
P_{D}\left( \left\langle t_{0},t\right\rangle \right) =P\left( E_{2}\cap
E_{1}\right) =P\left( E_{2}/E_{1}\right) $ $P\left( E_{1}\right) $ $,$ \ \ \
\ \ \ \ \ \ \ \ $(4)$

\bigskip

where \ $P\left( E_{1}\right) $ \ \ is the probability that a particle has
momentum \ $\overrightarrow{p}\in \Omega _{D}$ \ and \ \ $P\left(
E_{2}/E_{1}\right) $ \ \ means that if $\ E_{1}$ \ set in then \ $E_{2}$ \ \
set in with the probability \ \ $P\left( E_{2}/E_{1}\right) $ .

If we confine ourselves to non-relativistic particle and the time
development of \ \ $\varphi \left( \overrightarrow{x}-\overrightarrow{x}%
_{0}\right) $ \ \ is given by \ $\left( t>t_{0}\right) $

\bigskip

\ \ \ \ \ \ \ \ \ \ \ \ \ \ \ \ \ \ \ \ \ \ $\Psi \left( \overrightarrow{x}%
,t\right) =e^{-iH\left( t-t_{0}\right) }$ $\varphi \left( \overrightarrow{x}-%
\overrightarrow{x}_{0}\right) =\dint \dfrac{d^{3}\overrightarrow{p}}{\left(
2\pi \right) ^{3/2}}$ $C\left( \overrightarrow{p}\right) $ $e^{-iE\left(
t-t_{0}\right) +i\overrightarrow{p}\left( \overrightarrow{x}-\overrightarrow{%
x}_{0}\right) }=$

\bigskip

\bigskip

\ \ \ \ \ \ \ \ \ \ \ \ \ \ \ \ \ \ \ \ \ \ \ \ \ \ \ \ \ \ \ \ \ \ \ \ \ \
\ \ \ \ \ \ \ \ \ $=$ $\dint d\Omega \left( \overrightarrow{n}\right)
\dint\limits_{0}^{\infty }\dfrac{p^{2}dp}{\left( 2\pi \right) ^{3/2}}$ $%
C\left( p\overrightarrow{n}\right) $ $e^{-iE\left( t-t_{0}\right) +ip%
\overrightarrow{n}\left( \overrightarrow{x}-\overrightarrow{x}_{0}\right) }=$

\bigskip

\ \ \ \ \ \ \ \ \ \ \ \ \ \ \ \ \ \ \ \ \ \ \ \ \ \ \ \ \ 

\ \ \ \ \ \ \ \ \ \ \ \ \ \ \ \ \ \ \ \ \ \ \ \ \ \ \ \ \ \ \ \ \ \ \ \ \ \
\ \ \ \ \ \ \ \ \ $=$ $\dint d\Omega \left( \overrightarrow{n}\right) $ $%
\Psi ^{\left( \overrightarrow{n}\right) }\left( \overrightarrow{x},t\right) $
$,$

\bigskip

where we put \ $\overrightarrow{p}=\left| \overrightarrow{p}\right| 
\overrightarrow{n}=p\overrightarrow{n}$ \ \ 

\bigskip

\ \ \ \ \ \ \ \ \ \ \ \ \ \ \ \ \ \ \ \ \ \ \ \ \ \ \ \ \ \ \ \ \ \ \ $%
P\left( E_{1}\right) =\dint\limits_{\Omega _{D}}d\Omega \left( 
\overrightarrow{n}\right) \dint\limits_{0}^{\infty }p^{2}dp$ $\left| C\left(
p\overrightarrow{n}\right) \right| ^{2}$ \ \ \ \ \ \ \ \ \ $\ \ \ \ \ \ \ \
\ \ \ \ \ \ \ \ \ \ \ \ \ \ \ \ \ \ (5)$\ \ \ 

\bigskip

and we propose for \ $P\left( E_{2}/E_{1}\right) $ \ \ the following
expression

\bigskip

\bigskip

\ \ \ \ \ \ \ \ \ \ \ \ \ \ \ \ \ \ \ \ \ \ \ \ \ \ \ \ \ \ \ \ \ \ $P\left(
E_{2}/E_{1}\right) =\dfrac{\dint\limits_{t_{0}}^{t}dt^{\prime
}\dint\limits_{V_{D}}d^{3}\overrightarrow{x}\text{ }\left| \Psi _{D}\left( 
\overrightarrow{x},t^{\prime }\right) \right| ^{2}}{\dint\limits_{t_{0}}^{%
\infty }dt^{\prime }\dint\limits_{V_{D}}d^{3}\overrightarrow{x}\text{ }%
\left| \Psi _{D}\left( \overrightarrow{x},t\right) \right| ^{2}}$ $,$ \ \ \
\ \ \ \ \ \ \ \ \ \ \ $\ \ \ \ \ \ \ \ \ \ \ \ \ \ \ \ \ \ \ \ (6)$

\bigskip

where

\ \ \ \ \ \ \ \ \ \ \ \ \ \ \ \ \ \ \ \ \ \ \ \ \ \ \ \ \ \ \ \ \ \ $\Psi
_{D}\left( \overrightarrow{x},t\right) =\dint\limits_{\Omega _{D}}d\Omega
\left( \overrightarrow{n}\right) $ $\Psi ^{\left( \overrightarrow{n}\right)
}\left( \overrightarrow{x},t\right) $ $.$

\bigskip

As for the expression \ (6) \ we stand up for opinion that if particle has
momentum \ $\overrightarrow{p}\in \Omega _{D}$ \ \ with the density of
probability \ \ $\left| C\left( \overrightarrow{p}\right) \right| ^{2}$ \ \
then it can be described by the wave function \ \ $\Psi _{D}\left( 
\overrightarrow{x},t\right) $ \ \ and can occur in the detector in time \ $%
t^{\prime }$ \ \ with the probability density \ (with respect to time)

\bigskip

\ \ \ \ \ \ \ \ \ \ \ \ \ \ \ \ \ \ \ \ \ \ \ \ \ \ \ \ \ \ \ \ \ \ \ \ \ \
\ \ \ \ \ \ \ \ \ \ \ \ \ \ \ \ \ $\dint\limits_{V_{D}}d^{3}\overrightarrow{x%
}$ $\left| \Psi _{D}\left( \overrightarrow{x},t^{\prime }\right) \right|
^{2} $ $.$

\bigskip

It is evident that \ \ $P_{D}\left( \left\langle t_{0},t\right\rangle
\right) $ \ \ is not decreasing function and in the case \ \ $C\left( p%
\overrightarrow{n}\right) =C\left( p\right) $ \ \ we get for \ \ $%
P_{D}\left( \left\langle t_{0},\infty \right\rangle \right) $ \ \ the
expression

\bigskip

\ \ \ \ \ \ \ \ \ \ \ \ \ \ \ \ \ \ \ \ \ \ \ \ \ \ \ \ \ \ \ \ \ \ \ \ \ $%
P_{D}\left( \left\langle t_{0},\infty \right\rangle \right) =\dfrac{\Omega
_{D}}{4\pi }$

\bigskip as necessary.

\bigskip

\textbf{4. \ Arrival time}

\bigskip

If we reduce the detector to the point, say  \ $\overrightarrow{x}_{D}$ ,\ \
then we can write

\bigskip

\bigskip

\ \ \ \ \ \ \ \ \ \ \ \ \ \ \ \ \ \ \ \ \ \ \ \ \ \ \ \ \ \ \ \ \ \ \ \ $%
P\left( E_{2}/E_{1}\right) =\dfrac{\dint\limits_{t_{0}}^{t}dt^{\prime }\text{
}\left| \Psi ^{\left( \overrightarrow{n}_{D}\right) }\left( \overrightarrow{x%
}_{D},t^{\prime }\right) \right| ^{2}}{\dint\limits_{t_{0}}^{\infty
}dt^{\prime }\text{ }\left| \Psi ^{\left( \overrightarrow{n}_{D}\right)
}\left( \overrightarrow{x}_{D},t^{\prime }\right) \right| ^{2}}$ \ $,$

\bigskip

where

\ \ \ \ \ \ \ \ \ \ \ \ \ \ \ \ \ \ \ \ \ \ \ \ \ \ \ \ \ \ \ \ \ \ \ \ $%
\overrightarrow{n}_{D}=\dfrac{\overrightarrow{x}_{D}-\overrightarrow{x}_{0}}{%
\left| \overrightarrow{x}_{D}-\overrightarrow{x}_{0}\right| }$ $.$

\bigskip

The quantity

\ \ \ \ \ \ \ \ \ \ \ \ \ \ \ \ \ \ \ \ \ \ \ \ \ \ \ \ \ \ \ \ \ \ \ $%
\dfrac{dP_{D}\left( E_{2}/E_{1}\right) }{dt}$ $\symbol{126}$ \ $\left| \Psi
^{\left( \overrightarrow{n}_{D}\right) }\left( \overrightarrow{x}%
_{D},t\right) \right| ^{2}$

\bigskip

can be interpreted as the density \ (with respect to \ $t$ ) \ of
probability that a particle emitted at time \ $t_{0}$ \ \ entered the
detector at time \ $t$ . \ Now we can define the average value of the time \ 
$T=t-t_{0\text{ \ }}$ \ of the arrival to the point \ $\overrightarrow{x}%
_{D} $ \ \ by

\bigskip

\bigskip

\ \ \ \ \ \ \ \ \ \ \ \ \ \ \ \ \ \ \ \ \ \ \ \ \ \ \ \ \ \ \ \ \ \ $%
\left\langle t-t_{0}\right\rangle =\dfrac{\dint\limits_{t_{0}}^{\infty }dt%
\text{ }\left( t-t_{0}\right) \text{ }\left| \Psi ^{\left( \overrightarrow{n}%
_{D}\right) }\left( \overrightarrow{x}_{D},t\right) \right| ^{2}}{%
\dint\limits_{t_{0}}^{\infty }dt\text{ }\left| \Psi ^{\left( \overrightarrow{%
n}_{D}\right) }\left( \overrightarrow{x}_{D},t\right) \right| ^{2}}$ \ $.$ $%
\ \ \ \ \ \ \ \ \ \ \ \ \ \ \ \ \ \ \ \ \ \ \ \ \ \ \ \ (7)$

\bigskip

\bigskip

\textbf{5. \ Concluding remarks}

\bigskip

The presented particle detector model is too simple to be considered
realistic. Although it illustrates some basic features of particle detector
yet many questions have remainded open. For example, if a considered source
had emitted at time \ $t_{0}$ \ \ $N$ \ \ particles what is probability that
the detector will register \ $n$ \ \ of them ? \ This question is not
resolved also in the other more realistic models. At the present time we are
not able to state to a what extent our approach can be usefull for solving
problems we study.

\bigskip

\bigskip

\textbf{References}

\bigskip

[1] \ J.G.Muga, R.Sala and J.P.Palao, Superlattices Microsruct. 23 (1998)
833,

\ \ \ \ quant-ph/9801043.

[2] \ A.Ruschhaupt, \ quant-ph/9912060.

[3] \ J.Leon, J.Julve, P.Pitanga and F.J. de Urries, \ quant-ph/0002011.

[4] \ J.Leon, \ quant-ph/0008025.

[5] \ J.J. Halliwell, \ Prog.Theor.Phys. 102 (1999) 707, \ quant-ph/9805057.

[6] \ L.S.Schulman, \ Time's Arrows and Quantum Measurement, \ Cambridge
Univ.Press 1997.

[7] \ B.Gavean and L.S.Schulman, \ J.Stat.Phys. 58 (1990) 1209.

[8] \ L.S.Schulman, \ Ann.Phys. 212 (1991) 315.

\bigskip

\ \ 

\end{document}